\normalfont

\documentclass[10pt,conference,letterpaper]{IEEEtran}
\usepackage[pdftex]{graphicx}
\usepackage{graphicx} 
\usepackage[table]{xcolor}
\usepackage{booktabs}
\usepackage{tabularx}
\usepackage{balance}
\usepackage{makecell}
\usepackage{array}

\def\fixme#1{\bgroup \color{red}{[{#1}]}\egroup}

\IEEEoverridecommandlockouts

\begin{document}

\bstctlcite{bstctl:etal}

\title{Characterizing the Impact of Graph-Processing Workloads on Modern CPU's Cache Hierarchy}

\author{\IEEEauthorblockN{Alexandre Valentin Jamet\IEEEauthorrefmark{1}\IEEEauthorrefmark{2},
Lluc Alvarez\IEEEauthorrefmark{1}\IEEEauthorrefmark{2},
Marc Casas\IEEEauthorrefmark{1}\IEEEauthorrefmark{2}
}
\IEEEauthorblockA{\IEEEauthorrefmark{1}Barcelona Supercomputing Center, Barcelona, Spain}
\IEEEauthorblockA{\IEEEauthorrefmark{2}Universitat Polit\`ecnica de Catalunya, Barcelona, Spain}

E-mail: \{alexandre.jamet, lluc.alvarez, marc.casas\}@bsc.es\\
}

\maketitle


\begin{keywords}
cache management, cache bypassing, big data, graph processing, workload evaluation, irregular workloads, micro-architecture
\end{keywords}

\vspace{-4mm}
\section{Extended Abstract}
\label{sec:extended_abstract}

In recent years, graph-processing has become an essential class of workloads with applications in a rapidly growing number of fields. Graph-processing typically uses large input sets, often in multi-gigabyte scale, and data-dependent graph traversal methods exhibiting irregular memory access patterns. Recent work~\cite{8257937} demonstrates that, due to the highly irregular memory access patterns of data-dependent graph traversals, state-of-the-art graph-processing workloads spend up to 80~\% of the total execution time waiting for memory accesses to be served by the DRAM. The vast disparity between the Last Level Cache (LLC) and main memory latencies is a problem that has been addressed for years in computer architecture. One of the prevailing approaches when it comes to mitigating this performance gap between modern CPUs and DRAM is cache replacement policies.

In this work, we characterize the challenges drawn by graph-processing workloads and evaluate the most relevant cache replacement policies.

\subsection{Graph-processing Workloads}
\label{subsec:graph_processing_workloads}

Graph-processing is a class of emerging workloads that, nowadays, can be found in various applications. Graph-processing can be found in both industry and academia, from social network analytics to web search engines and biomedical applications.

Graph-processing typically uses sparse data formats such as \textit{Compressed Sparse Row/Column (CSR/CSC)} to manage a large amount of data. The CSR/CSC format is used to encode the graph adjacency matrix using two data structures: 1) the \textit{Offset Array (OA)} and; 2) the \textit{Neighbours Array (NA)}. Finally, additional data structures contain numerical data corresponding to a graph's vertices called \textit{Property Arrays (PA)}.

Manipulating these sparse data structures often produces irregular memory access patterns. For example, when computing the \textit{Sparse Matrix-Vector (SpMV)} multiplication $y = A \cdot x$, accesses to vector $x$ are indexed by the column indices of matrix $A$, which are non-contiguous and constitute an irregular access stream. Graph-processing workloads also display highly irregular memory access patterns driven by operations like graph traversals that  require visiting all the vertices $V$ of a graph, that is, scanning the adjacency matrix rows following the graph's connectivity.

\begin{figure}[h]
    \centering
    \includegraphics[width=\columnwidth]{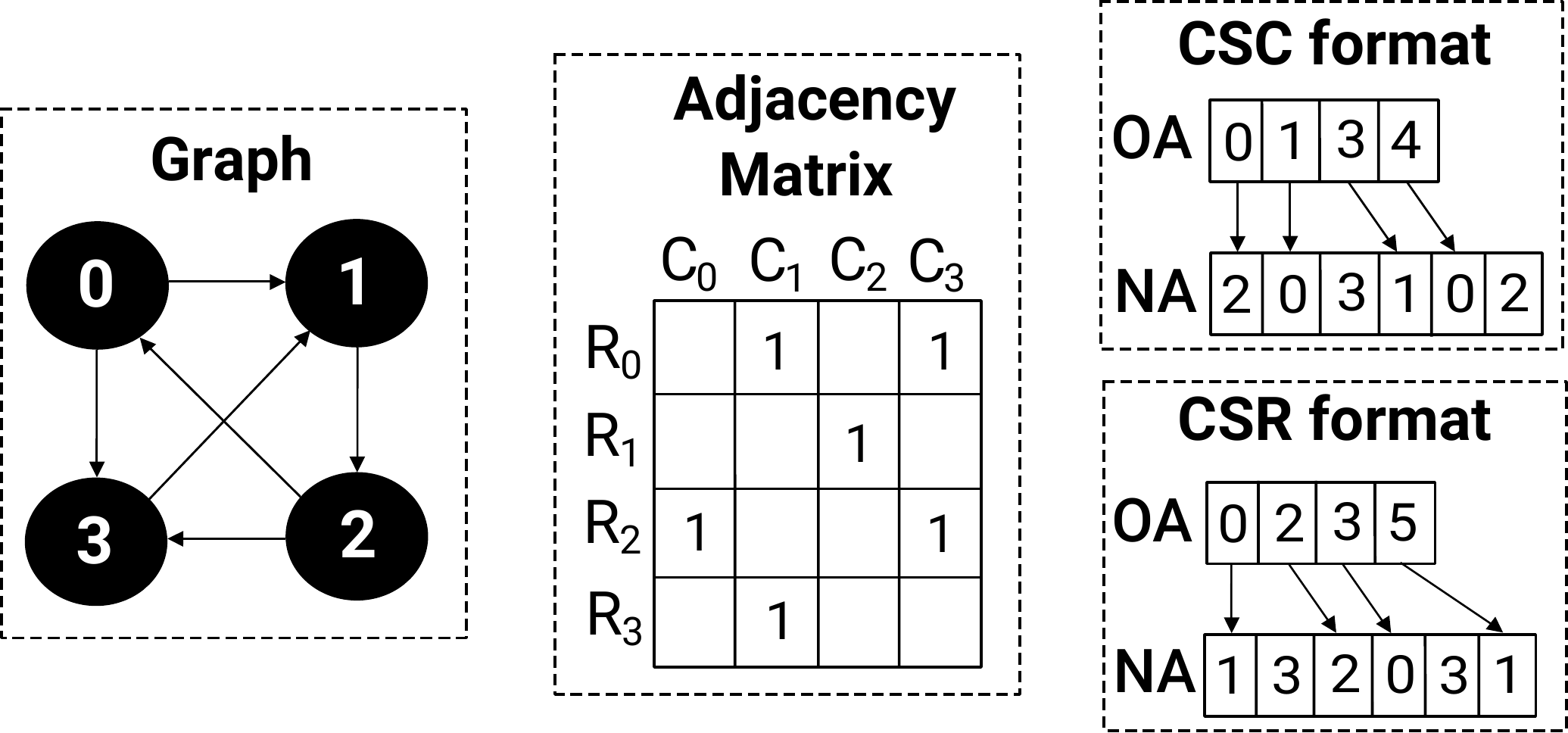}
    \caption{Example of a graph representation in memory using the CSR/CSC formats.}
    \label{fig:graph_traversal}
\end{figure}

Figure~\ref{fig:graph_traversal} shows an example graph along with its adjacency matrix and its representation using either the CSR or the CSC formats.

\subsection{Cache Replacement Policies}
\label{subsec:cache_replacment_policies}

For this work, we evaluate six of the most relevant replacement policies for the LLC. SRRIP, DRRIP~\cite{jaleel_high_2010} hahandleshe replacement process by predicting reuse distances. SHiP~\cite{wu_ship:_2011} leverages SRRIP and extends it with the addition of a PC feature. Hawkeye~\cite{jain_back_2016}, Glider~\cite{shi_applying_2019} a,nd MPPPB~\cite{jimenez_multiperspective_2017} further improve by making the addition of machine learning inspired techniques (\textit{e.g.}: perceptron, SVM, etc.). Specifically, these advanced cache replacement policies leverage micro-architectural features based on bits extracted from the program counters and virtual/physical addresses to establish correlations and produce predictions.

\subsection{Experimental Setup}

Our evaluation considers ChampSim, a detailed trace-based simulator that models a Cascade Lake micro-architecture. The micro-architecture simulated has only one core, L1 instruction, and data cache of 32KB each, an L2 cache of 1MB, and an L3 cache of 1.375MB. The system also includes an 8GB main memory based on DDR4 SDRAM with a data rate of 2.933GT/s.

\subsection{Experimental results}

\begin{figure}[h]
    \centering
    \includegraphics[width=\columnwidth]{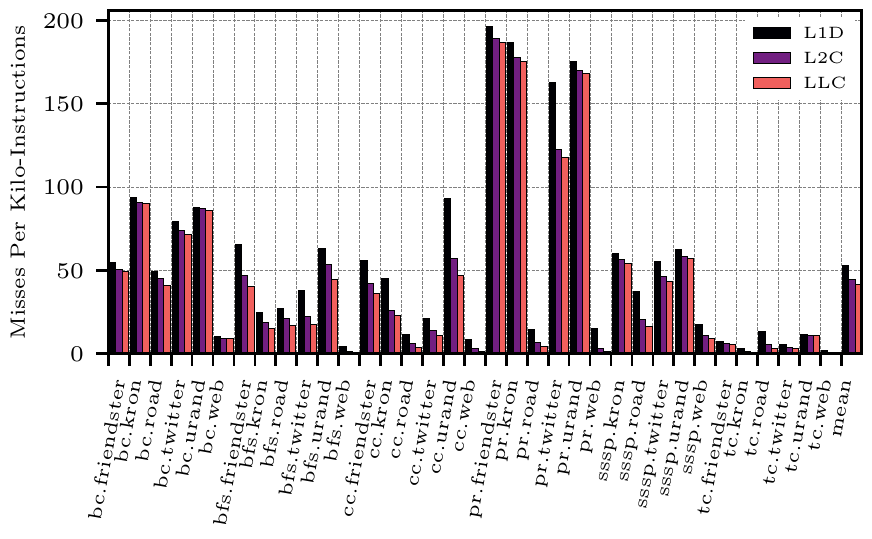}
    \caption{Misses-Per-Kilo-Instruction (MPKI) across the different levels of the cache hierarchy triggered by graph-processing workloads.}
    \label{fig:gapbs_mpki_all_caches}
\end{figure}

Figure~\ref{fig:gapbs_mpki_all_caches} shows the MPKI rates in all three cache hierarchy levels for workloads of the GAP~\cite{beamer_gap_2017} benchmark suite. This figure shows that graph-processing workloads suffer from a large number of misses at all levels of the cache hierarchy. The average MPKI rates of these workloads in the L1D, L2C, and LLC are 53.2, 44.2 and 41.8. We can further observe that a considerable portion (78.6\%) of the accesses that trigger L1D misses also miss in the lower levels of the cache hierarchy and require a DRAM access.

Cache replacement policy is an obvious topic when it comes to improving the behavior of the cache hierarchy for certain types of workloads. As such, we evaluated the replacement policies presented in \ref{subsec:cache_replacment_policies} to understand their impact on performance.

\begin{figure}[h]
    \centering
    \includegraphics[width=\columnwidth]{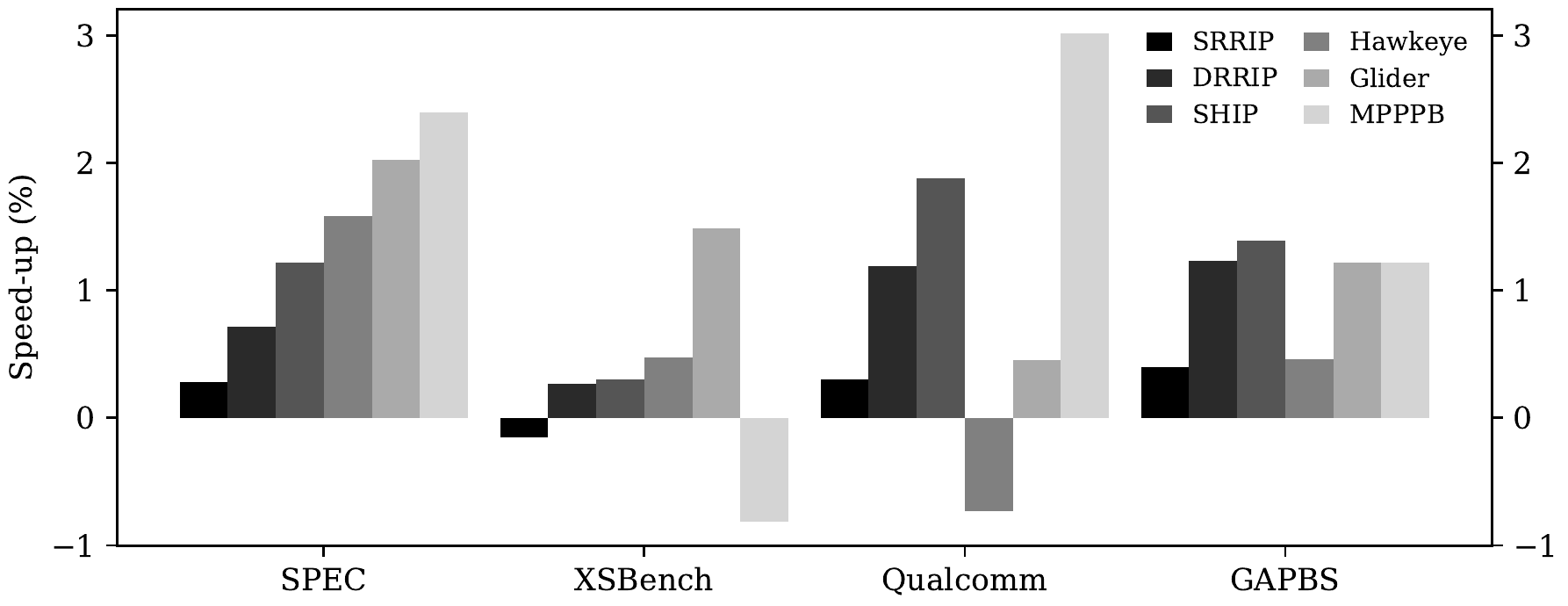}
    \caption{Geometric mean speed-up over LRU of state-of-the-art LLC replacement policies for the different benchmark suites.}
    \label{fig:speedup_split_suites}
\end{figure}

Figure~\ref{fig:speedup_split_suites} shows the geometric mean speed-up of the state-of-the-art cache replacement policies evaluated over the baseline LRU policy for various benchmark suites, including SPEC 2006 \& 2017, along with the GAP workloads. The results show that the different policies can catch different kinds of access patterns and benefit different workloads.

These results show that more complex replacement policies such as Hawkeye, Glider, and MPPPB have difficulties generalizing to benchmark suites beyond SPEC 2006 \& 2017. This is due to the underlying assumptions on memory access patterns used to build these complex cache replacement policies. As shown in~\ref{subsec:graph_processing_workloads}, graph-processing is a prime example where the number of PC is very limited and where each PC maps to a very large number of addresses making correlations nearly impossible to establish.

\subsection{Conclusion}

Overall, this work highlights the poor ability of state-of-the-art cache replacement policies to leverage significant benefits against a baseline using an LRU policy for graph-processing workloads, despite the very high hardware complexity of such techniques. We show pieces of evidence that this bleak outlook stems from two factors: i) the very distinct nature of graph-processing workloads and; ii) the immense pressure these workloads create on the cache hierarchy.  


\IEEEpeerreviewmaketitle

\vspace{-2mm}

\section{Acknowledgment}
This work has been published in Proceedings of the International Symposium on Workload Characterization (IISWC), 2020 \cite{jamet_characterizing_2020}.  

\vspace{-2mm}

%
\balance
\bibliographystyle{IEEEtran}
\bibliography{IEEEabrv,paper}
\vspace{-7mm}

\begin{biography}[
{
\includegraphics[width=1in,height=1.25in,clip,keepaspectratio]{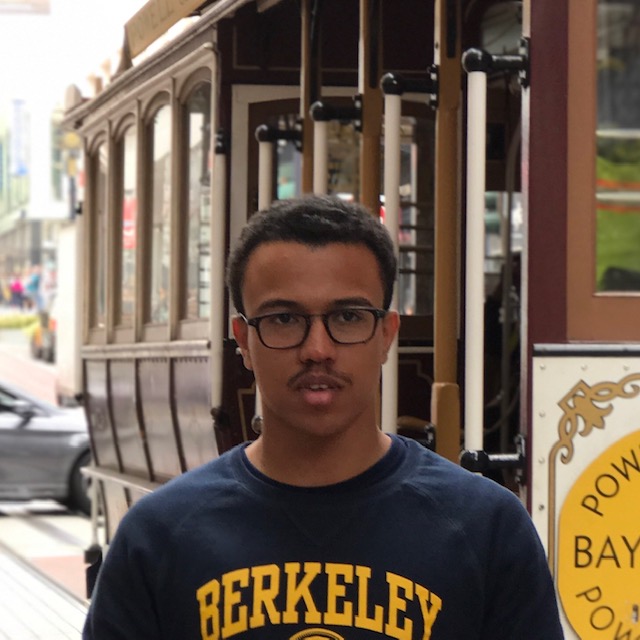}
}
]
{Alexandre Valentin Jamet} studied two years of Higher School Preparatory Classes with a Physics and Engineering Sciences major at LGT Baimbridge, Guadeloupe. In the following years, he pursued his MSc degree in parallel with an Engineer Diploma from TELECOM Nancy with a major in Embedded Computing. He concluded his studies in Nancy in 2018. Since 2018, he has been a Ph.D. candidate at the Computer Architecture departments of Barcelona Supercomputing Center (BSC) and Universitat Polit\`ecnica de Catalunya (UPC), Spain.

\end{biography}

\end{document}